\documentclass[nofootinbib,superscriptaddress,notitlepage, 10pt,twocolumn]{revtex4-1}
\usepackage{graphicx}
\usepackage[colorlinks=true,citecolor=red,urlcolor=black]{hyperref}
\usepackage{amsfonts}
\usepackage{amsmath,amssymb}
\usepackage{dsfont}
\usepackage{euscript}
\usepackage{float}
\usepackage{amsthm}
\usepackage{mathalfa}
\usepackage{bbm}
\usepackage{ragged2e}
\usepackage{color}
\usepackage[dvipsnames]{xcolor}
\usepackage[normalem]{ulem}

\usepackage{xparse}
\usepackage{framed}
\usepackage{tikz}
\usetikzlibrary{matrix,backgrounds}
\usepackage{float}
\def\be{\begin{equation}}
\def\ee{\end{equation}}

\newcommand{\Pmat}{\textbf{p}}
\newcommand{\Gmat}{{G}}
\newcommand{\eps}{\varepsilon}

\newcommand{\gell}{{G}^\ell}
\newcommand{\ginf}{{G}^\infty}
\newcommand{\la}{\langle}
\newcommand{\ra}{\rangle}
\def\be{\begin{equation}}
\def\ee{\end{equation}}
\def\bi{\begin{itemize}}
\def\ei{\end{itemize}}

\newcommand{\da}{\dagger}
\newcommand{\tE}{\tilde{E}}
\newcommand{\piax}{E^a_x}
\newcommand{\piby}{E^b_y}
 \newcommand{\ketz}{\ket{\phi_z}}
\newcommand{\braz}{\bra{\phi_z}}
 \newcommand{\ketzp}{\ket{\phi_{z'}}}

\newcommand{\qset}{\mathcal{Q}(\lambda)}
\newcommand{\Sset}{\mathcal{S}}

\newcommand{\sket}[1]{{\ensuremath{\lvert#1\rangle}}}
\newcommand{\lket}[1]{{\ensuremath{\left\lvert#1\right\rangle}}}
\newcommand{\ket}[1]{\if@display\lket{#1}\else\sket{#1}\fi}

\newcommand{\sbra}[1]{{\ensuremath{\langle#1\rvert}}}
\newcommand{\lbra}[1]{{\ensuremath{\left\langle#1\right\rvert}}}
\newcommand{\bra}[1]{\if@display\lbra{#1}\else\sbra{#1}\fi}

\newcommand{\sbraket}[2]{{\ensuremath{\langle#1\rvert#2\rangle}}}
\newcommand{\lbraket}[2]{{\ensuremath{\left\langle#1\!\left\rvert\vphantom{#1}#2\right.\!\right\rangle}}}
\newcommand{\braket}[2]{\if@display\lbraket{#1}{#2}\else\sbraket{#1}{#2}\fi}

\newcommand{\sketbra}[2]{{\ensuremath{\lvert #1\rangle\!\langle #2\rvert}}}
\newcommand{\lketbra}[2]{{\ensuremath{\left\lvert #1\right\rangle\!\!\left\langle #2\right\rvert}}}
\newcommand{\ketbra}[2]{\if@display\lketbra{#1}{#2}\else\sketbra{#1}{#2}\fi}

\newcommand{\proj}[1]{\ketbra{#1}{#1}}

\begin{document}
\title{Characterising the correlations of prepare-and-measure quantum networks  }

\author{Yukun Wang}
\affiliation{Department of Electrical \& Computer Engineering, National University of Singapore, Singapore }
\author{Ignatius William Primaatmaja}
\affiliation{Centre for Quantum Technologies, National University of Singapore, Singapore}
\author{Emilien Lavie}
\affiliation{Centre for Quantum Technologies, National University of Singapore, Singapore}
\affiliation{T\'{e}l\'{e}com ParisTech, LTCI, Paris, France }
\author{Antonios Varvitsiotis}
\affiliation{Department of Electrical \& Computer Engineering, National University of Singapore, Singapore }
\author{Charles Ci Wen \surname{Lim}}
\email{charles.lim@nus.edu.sg}
\affiliation{Department of Electrical \& Computer Engineering, National University of Singapore, Singapore }
\affiliation{Centre for Quantum Technologies, National University of Singapore, Singapore}

\begin{abstract} Prepare-and-measure (P\&M) quantum networks are the basic building blocks of quantum communication and cryptography. These networks crucially rely on non-orthogonal quantum encodings to distribute quantum correlations, thus enabling superior communication rates and information-theoretic security. Here, we present a computational toolbox that is able to efficiently characterise the set of input-output probability distributions for any discrete-variable P\&M quantum network, assuming only the inner-product information of the quantum encodings. Our toolbox is thus highly versatile and can be used to analyse a wide range of quantum network protocols, including those that employ infinite-dimensional quantum code states. To demonstrate the feasibility and efficacy of our toolbox, we use it to reveal new results in multipartite quantum distributed computing and quantum cryptography. Taken together, these findings suggest that our method may have implications for quantum network information theory and the development of new quantum technologies.    \end{abstract}

\maketitle

 \bigskip
\noindent \textbf{Introduction.~}Quantum correlations~\cite{Horodecki2009,Bell1964,Wiseman2007} (namely, entanglement, nonlocality, steering correlations, etc) are essential resources in quantum information processing. In short, they are the reason why we see such unique advantages in quantum communication, cryptography, computing, and imaging. The general observation is that the stronger these correlations are, the more powerful quantum information becomes. This is especially the case for quantum communication~\cite{Gisin2007} and quantum cryptography\cite{Gisin2002}, where stronger entanglement means higher quantum fidelity and stronger information security. For this reason, the characterisation of quantum correlations is an integral step in many quantum information protocols and a central research topic in quantum information science.

In this work, we are interested in characterising the correlations of prepare-and-measure (P\&M) quantum networks. These are the basic building blocks of quantum communication and quantum cryptography. The central task of a P\&M quantum network is to send a classical message $z$ over a quantum network to a group of receivers. This message could be anything, e.g., a secret key, elements of a database, or a signed certificate---it depends on the function of the protocol. Quantum encoding is done by preparing a quantum signal in one of the $n$ predefined pure states, $\{\ket{\psi_z}\}_{z=1}^n$ (determined by the input $z$), and decoding is accomplished by making a measurement (sampled from a finite set of decoding settings) on the output quantum signal. For a generic P\&M quantum network with $k$ spatially separated receivers, we write $p(a_1a_2\ldots a_m|x_1x_2\ldots x_k, z)$ to denote the probability of obtaining outcomes $a_1a_2\ldots a_k$ given decoding functions $x_1x_2\ldots x_k$ and message $z$. We further use $\Pmat$ to denote the entire list of input-output probability distributions. 

Our broad goal is to reveal the fundamental limits of P\&M quantum networks without restrictions on the network and local decoding strategies. In particular, we are interested in identifying the set of quantum-realisable correlations $\Pmat$ (henceforth called the quantum set) using only the knowledge of the quantum encoding scheme $\{\ket{\psi_z}\}_z$ as the constraining factor. This type of approach is extremely useful for analysing the performance of quantum communication and quantum cryptography. For instance, one can use the quantum set to derive lower bounds on the quantum network's error probabilities~\cite{Yard2011,Hirche2015,Savov2015}. These bounds essentially tell us what the quantum encoding $\{\ket{\psi_z}\}_z$ could achieve in practice, be it for quantum cryptography, communication, or distributed computing purposes, as we shall show later.

Also, from the perspective of quantum information theory, this approach draws a direct connection between the distinguishability of quantum states and quantum correlations. More concretely, we first note that if the quantum encoding $\{\ket{\psi_z}\}_z$ is completely orthogonal, i.e., $\braket{\psi_z}{\psi_{z'}}=\delta_{zz'}$, then $\Pmat$ is generally `unconstrained'. That is, such encodings are classical states and hence can be arbitrarily copied--- as such, there are no physical principles that constrain what the input-output probability distribution could be (except for the usual normalisation requirements). The interesting part comes when the encoding $\{\ket{\psi_z}\}_z$ is non-orthogonal. In this case, there are two unique consequences. Firstly, it is generally impossible for every receiver to learn the same information about $z$. This is due to the fact that one cannot clone non-orthogonal states~\cite{Wootters1982}, and consequently, there is a global trade-off between the amount of accessible information that each receiver can receive~\cite{Fuchs1996,Horodecki2005}. Secondly, no receiver can completely learn $z$ even if he or she has received $\ket{\psi_z}$ with perfect fidelity. This is because non-orthogonal states are fundamentally indistinguishable: there is no measurement that can discriminate them with perfect reliability~\cite{Holevo1973}. Consequently, probability assignments like $p(a=z|x)=1$ are forbidden. Taken together, these imply that, contrary to orthogonal (classical) encodings, correlations emanating from quantum encodings have non-trivial constraints (e.g., see quantum broadcasting~\cite{Barnum1996,Barnum2007}).

 \bigskip
\noindent \textbf{Results.~}To tackle the above characterisation problem, we propose a general computational method that is able to approximate (from the outside) the quantum set of any P\&M quantum network. The approximation is based on a hierarchy of semidefinite relaxations, which is a generalisation and novel application of earlier research in quantum nonlocality~\cite{Bell1964,Brunner2014,Tsirelson1987,Landau1988,Wehner2006, Miguel2007,Miguel2008}. 
A key feature of our method is that it is semi-device-independent (SDI)~\cite{SDI2011,Bowles2014, Lunghi2015, Woodhead2015, Berta2015, Himbeeck2017, Brask2017}. That is, the analysis provided is independent of how the network and measurements are implemented (as required above). The method only requires that the quantum encoding $\{\ket{\psi_z}\}_z$ is characterised in terms of its Gram matrix, i.e., $\braket{\psi_z}{\psi_{z'}}=\lambda_{zz'}$, which in practice can be easily obtained by taking the inner products of the quantum code states (i.e., using their specifications). 

The main advantage with this approach is that the dimension of the encoding system is no longer necessary in the analysis: using the Gram matrix information alone is enough to characterise the quantum set. In this sense, our approach is more practical than the standard SDI approach, which assumes the dimension of the quantum encoding system~\cite{SDI2011,Bowles2014, Lunghi2015, Woodhead2015}; notice that physical dimension is generally difficult to fix in practice as actual systems have multiple degrees of freedom. We remark that alternative SDI approaches based on bounded energy constraints~\cite{Himbeeck2017} and the transmission of non-orthogonal binary states~\cite{Brask2017} have also been proposed. These have similar advantages as our approach, but present analyses using these approaches are so far limited to binary code states. It remains to be seen if these can be readily generalised to multiple code states, schemes which are often used in quantum technologies. In the following, we show that our method can be used to efficiently analyse any practical quantum communication protocol~\cite{Arrazola2014}, including those that use multiple infinite-dimensional code states.

\begin{figure*}[t!]
\includegraphics[width=11cm]{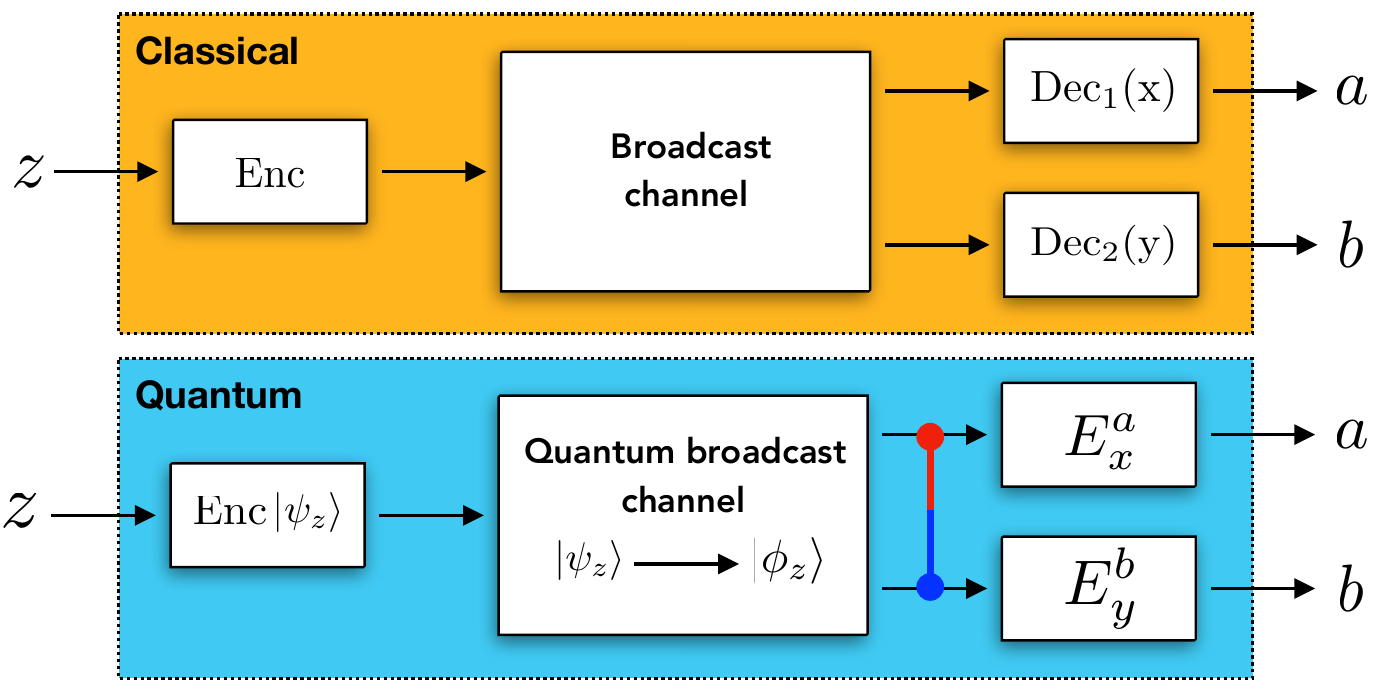}\caption{\footnotesize{\textbf{Scenario and assumptions.}~The behaviour of a two-receiver P\&M quantum network is generally described by $\Pmat=[p(ab|xy,z)]$, which expresses the probability of $z$ transiting to outcomes $a,b$ given measurement inputs $x,y$. In the quantum setting, the set of conditional probabilities are given by $p(ab|xy, z)={\bra{\phi_z} \piax\piby\ket{\phi_z}}$, with the constraint that $\braket{\phi_z}{\phi_{z'}}=\lambda_{zz'}$ is fixed. Our consideration hence assumes three conditions: (1) the set of code states are pure states, (2) the Gram matrix of these states is known, and (3) the receivers are independent of each other (they do not share any quantum resources, although classical randomness is allowed).     }    }\label{fig1}
\end{figure*}

To keep our presentation concise, we restrict the discussion to two-receiver P\&M quantum networks (see Figure~\ref{fig1}); extension to larger networks is straightforward. Consider a prepare-and-measure quantum task, where random code states are sent across a network to two independent receivers, called Alice and Bob, for measurement. For simplicity, we divide the task into two phases: a distribution phase and a measurement phase. In the first phase, a classical random source $z$ is encoded into a quantum system $\ket{\psi_z}$ and distributed to Alice and Bob via an untrusted quantum network. For this type of transmission, it is useful to work in the purification picture, where state transformations are given by unitary evolutions~\cite{Wilde2013}. That is, by working in a higher-dimensional Hilbert space, we may see the transmission as an isometric evolution that takes $\ket{\psi_z}$  to some pure output state $\ket{\phi_z}$, which is now shared between the receivers and the network environment (the purification system). The key advantage of this picture is that while the dimension and possibly other properties of $\ket{\psi_z}$ may change after the transmission, the inner-product information of $\{\ket{\phi_z}\}_z$ remains the same: $\braket{\phi_z}{\phi_{z'} }=\braket{\psi_z}{\psi_{z'}}$. Importantly, this means that our initial knowledge about $\braket{\psi_z}{\psi_{z'}}=\lambda_{zz'}$ is preserved in the transformed states.

In the measurement phase, Alice and Bob perform independent and random measurements on $\ket{\phi_z}$ to gain information about $z$. Since there are only two receivers here, we revert back to the usual convention and denote Alice's and Bob's measurements by $x$ and $y$ and their corresponding measurement outcomes by $a$ and $b$, respectively. Then, using the quantum Born rule, we have that the probability of observing outcomes $a,b$ given measurements $x, y$ and $\ket{\phi_z}$ is
\be \label{eq1:quantum}
p(ab|xy, z)={\bra{\phi_z} \piax\piby\ket{\phi_z}},
\ee
where $\{\piax\}$ and $\{\piby\}$ are projective measurements satisfying the following properties: (\emph{i}) for any $x$, $\piax E^{a'}_x=0$ for $a\not=a'$, (\emph{ii}) $\sum_a \piax=\mathbb{I} $, (\emph{iii}) $(\piax)^2=\piax=(E^a_x)^{\dagger}$, and (\emph{iv}) $[\piax,\piby]=0$. We note that there is no loss of generality in assuming projective measurements here. Indeed, we can always lift any measurement to a projective one by working in a higher-dimensional Hilbert space; in our case this is possible since the dimension of the network is not fixed.  The last property reflects the fact that Alice's and Bob's measurements are separable and hence the application of one has no effect on the outcome of the other.

Our characterisation problem is thus the following: Given an $n\times n$ Hermitian positive-semidefinite matrix  $\lambda$, what is the corresponding quantum set~$\Pmat$? We denote this set by $\qset$. In principle, solving this problem would require optimising over all possible quantum states and measurements in equation (\ref{eq1:quantum}) subject to the constraints $\braket{\phi_z}{\phi_{z'}}=\lambda_{zz'}$. However, this task is computationally intractable: the dimension of the network is not fixed and thus could be infinite. To overcome this obstacle, we take inspiration from the characterisation techniques~\cite{Tsirelson1987,Landau1988,Wehner2006, Miguel2007,Miguel2008} developed in Bell nonlocality research~\cite{Bell1964,Brunner2014}, which is a special case of our problem. Recall that in a Bell experiment, local random measurements are made on a fixed source ${\ket{\phi}}$ instead of a varying source ${\ket{\phi_z}}$. Notably, it was shown in refs~\cite{Miguel2007,Miguel2008} that the set of quantum probabilities derived from Bell experiments can be approximated via a hierarchy of membership tests. There, the basic idea is to bound the quantum set using a sequence of weaker (but tractable) characterisation tasks, which nevertheless still represent very well the original problem.

{In this work}, we show that a similar characterisation technique can also be devised for the general problem. More specifically, we give a general procedure for deriving (tractable) necessary conditions for any discrete-variable P\&M quantum network. To start with, consider a quantum probability distribution $p(ab|xy, z)={\bra{\phi_z} \piax\piby\ket{\phi_z}}$, where ${\braket{\phi_z}{\phi_{z'}}}=\lambda_{zz'}$, and with $\{\piax, \piby\}_{a,x,b,y}$ satisfying properties~{(\emph{i})--(\emph{iv})}. Let $\Sset=\{S_1,\ldots,S_m\}$ be a finite set of $m$ operators, where each element is a linear combination of products of $\{\piax, \piby\}_{a,x,b,y}$.
Then define $\Gmat$ to be the $nm\times nm$ block matrix
\[
\Gmat=\sum_{z,z'=1}^nG^{zz'}\otimes \ketbra{e_z}{e_{z'}},
\]
where $G^{zz'}_{(i,j)}=\braz S_i^\dag S_j\ketzp$ for all $z,z'\in [n], i,j\in [m]$.
Here we denote by $\{\ket{e_z}\}_{z=1}^n$ the standard orthonormal basis  of $\mathbb{C}^n$ and by $G^{zz'}_{(i,j)}$ the $ij$-entry of the matrix~$G^{zz'}.$ By construction, the matrix $\Gmat$ is Hermitian and positive-semidefinite (PSD)~\cite{Horn2013}. Furthermore,   properties (\emph{i})--(\emph{iv}) of the measurement operators  and the inner-product constraints $\braket{\phi_z}{\phi_{z'}}=\lambda_{zz'}$ translate to linear conditions on the entries of $\Gmat$. To see this, we note that if the set  $\Sset$ contains operators $\{\piax\}_{a,x} $ and $\{\piby\}_{b,y} $, then it can be easily verified that $\Gmat$ satisfies
\begin{eqnarray*}
\sum_{b}G^{zz}_{(a,x),(b,y)} &=&\sum_b p(ab|xy,z) \\  \sum_{a}G^{zz'}_{(a,x),(a,x)}&=&\lambda_{zz'}.
\end{eqnarray*}

Therefore, for any discretely modulated P\&M quantum network, it is always possible to define a PSD matrix that captures the original quantum model (\ref{eq1:quantum}) in terms of constraints that are linear in its entries. Importantly, the existence of such a matrix provides us with a powerful means to check if a given $\Pmat$ is of quantum origin. More specifically, we can use semidefinite programming (SDP) techniques~\cite{Boyd1996} to verify if $\Pmat$ is in the set of compatible PSD matrices: if $\Pmat$ is not a member, we conclude that it is not quantum realisable. However, successful membership does not necessarily mean $\Pmat$ is of quantum origin. This is due to the fact that our characterisation method is a semidefinite relaxation~\cite{Burgdorf2016} of the original problem and hence can only provide an outer-approximation of $\qset$. 

However, by introducing additional linear constraints via a hierarchical procedure, it is possible to gain a tighter characterisation of $\qset$. In particular, we could use the hierarchy proposed in refs.~\cite{Miguel2007,Miguel2008} to build a series of increasingly stringent membership tests, where the associated Gram matrix $\Gmat$ grows bigger in each step and more constraints are generated. More precisely, we define a sequence of hierarchical sets $\Sset_1=\{\piax, \piby \}$, $\Sset_2=\Sset_1\cup\{E^a_xE^{a'}_{x'}\} \cup \{E^b_yE^{b'}_{y'} \}\cup \{E^a_xE^b_y\},$ where $\Sset_{k}$ is defined inductively as the set of all operator sequences constructed from $\piax, \piby$ satisfying $\Sset_k \subseteq \Sset_{k+1}$. This corresponds to a sequence of Gram matrices, $\Gmat^1,\Gmat^2,\ldots$ with increasing size and constraints. Since the Gram matrix $\Gmat^k$ of a particular $k$th step is at least as informative as a smaller sized Gram matrix $\Gmat^{k'}$, we conclude that the approximated set $\qset_{k}$ is a subset of $\qset_{k'}$. Therefore, moving up the hierarchy gives a tighter approximation of the quantum set: $\qset \subseteq \qset_k \subseteq \qset_{k-1}\ldots$. In Appendix~\ref{App:A}, we prove that this hierarchy is in fact sufficient: it converges to the quantum set, $\lim_{k\rightarrow \infty}\qset_k=\qset$. Nevertheless, in the applications below, we see that low-level approximations are already enough to achieve very tight bounds.

 \bigskip
\noindent \textbf{Applications.~}Our method can be applied to any quantum communication task that employs the prepare-and-measure scheme. To illustrate this point, we provide two examples of application: (1) distributed quantum random access coding (QRAC)~\cite{Ambainis1999,Nayak1999} and (2) quantum key distribution (QKD)~\cite{Scarani2009, Lo2014}.

In the first, we consider a distributed computing task where two random bits $z_0z_1$ are encoded into a quantum state $\ket{\psi_{z_0z_1}}$ and sent to Alice and Bob for selective decoding. For the decoding part, Alice and Bob are each given a random position bit and their goal is to guess the input bit that is associated with the position bit. For example, if Alice receives $x=1$, she has to guess the value of $z_1$ via measurement on her share of $\ket{\psi_{z_0z_1}}$. This task can be seen as a type of distributed quantum database, where network users can choose to learn any entry of the database; this includes the case whereby multiple users can choose to learn the same entry. To this end, we quantify the network's ability to distribute information by Alice's and Bob's guessing probabilities, which we denote by $p(a=z_x)$ and $p(b=z_y)$, respectively.

At this point, it is useful to recall that if $\ket{\psi_{z_0z_1}}$ is a two-level quantum system (i.e., a qubit), then the best encoding strategy (in the case of the standard two-party QRAC) is to use the so-called conjugate coding scheme~\cite{Wiesner1983}: $\ket{\psi_{00}}=\ket{+}$, $\ket{\psi_{10}}=\ket{+i}$, $\ket{\psi_{01}}=\ket{-i}$, and $\ket{\psi_{11}}=\ket{-}$, where $\ket{\pm}$ and $\ket{\pm i}$ are the eigenstates of the Pauli operators $\mathbb{X}$ and $\mathbb{Y}$, respectively. This gives a guessing probability of $\left(1+1/\sqrt{2}\right)/2 \approx 0.853$~\cite{Ambainis1999,Nayak1999}, which is optimal for qubit code states. Interestingly, using our method, we find that similar bounds can be established using only the Gram matrix information of the code states. In particular, we consider a set of code states $\{\ket{\psi_{00}},\ket{\psi_{11}},\ket{\psi_{10}},\ket{\psi_{01}}\}$, whose Gram matrix is fixed to that of $\{\ket{+},\ket{-},\ket{+i},\ket{-i}\}$, and ask what is Alice's optimal guessing probability given Bob's guessing probability is fixed. Our method predicts the following quantum boundary: $(2p(a=z_x)-1)^2+(2p(b=z_y)-1)^2\leq 1/2$, which is drawn in Figure \ref{fig2}. The semidefinite program for this optimisation is provided in Appendix~\ref{App:B}. 

Three remarks are in order here. Firstly, we see that the boundary (obtained from $\qset_1$) gives the same upper bound as ref.~\cite{Ambainis1999} when one of the receivers is restricted to random guessing. This can be seen as the case in which one party receives $\ket{\psi_{z_0z_1}}$ with perfect fidelity and the other party receives a dummy state. Secondly, the boundary specifies a non-trivial trade-off function between Alice's and Bob's guessing probabilities, which is independent of their measurement strategies. This implies that the bound is absolute and cannot be improved upon with better measurement strategies, even if Alice and Bob are allowed to use shared randomness. Thirdly, although our method can only provide an outer-approximation of the quantum set, it turns out that the first level of the hierarchy is already tight. More specifically, there is a concrete example which saturates the boundary predicted by $\qset_1$; see Figure \ref{fig2} for more details. This example is given by the optimal asymmetric qubit cloning machine~\cite{Cerf1998}, which optimally splits the qubit information between multiple parties (according to some predefined ratio); this is indeed a natural choice as the goal of the network is to preserve as much quantum information as possible for each party while splitting it. 

\begin{figure}[t]
\includegraphics[width=8.4cm]{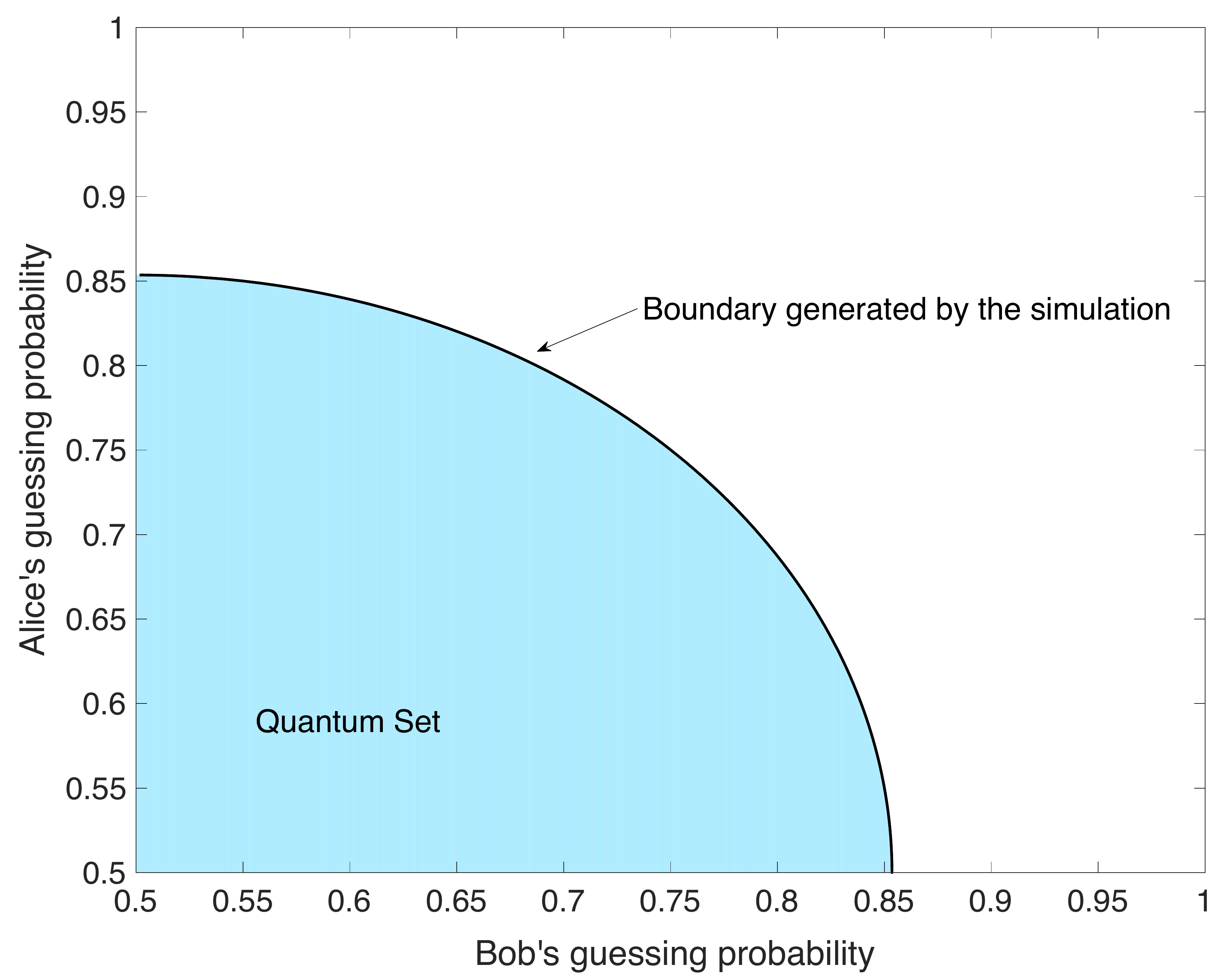}\caption{\footnotesize{\textbf{Distributed two-receiver QRAC.} The boundary is generated using the first level of the SDP hierarchy. In principle, the boundary is not necessarily tight, for $\mathcal{Q}(\lambda)_1$ may contain correlations that are not of quantum origin. However, in our case we show that the derived boundary is tight: it is saturated by the optimal asymmetric qubit cloning machine. Suppose the quantum code states are given by the conjugate coding scheme: $\ket{\psi_{00}}=\ket{+}\ket{0}$, $\ket{\psi_{10}}=\ket{+i}\ket{0}$, $\ket{\psi_{01}}=\ket{-i}\ket{0}$, and $\ket{\psi_{11}}=\ket{-}\ket{0}$. The quantum network is assumed to be an asymmetric cloning channel $U_f$: $\ket{0}\ket{0} \rightarrow \ket{0}\ket{0}$ and $\ket{1}\ket{0} \rightarrow \sqrt{1-f}\ket{1}\ket{0}+\sqrt{f}\ket{0}\ket{1}$, where $f \in [0,1]$. For the decoding, we assume that Alice and Bob perform the optimal QRAC qubit measurements: $\piax=(\mathbb{I}+(-1)^a(\mathbb{X}+(-1)^{x+1}\mathbb{Y})/\sqrt{2})/2 $, $\piby=(\mathbb{I}+(-1)^b(\mathbb{X}+(-1)^{y+1}\mathbb{Y})/\sqrt{2})/2$. Using these and setting the left subsystem as Alice's and the right subsystem as Bob's, we have that $p(a=z_x)=1/2+\sqrt{(1-f)/2}/2$ and $p(b=z_y)=1/2+\sqrt{f/2}/2$. These give $(2p(a=z_x)-1)^2+(2p(b=z_y)-1)^2= 1/2$, which is the quantum boundary predicted by the $\qset_1$ set.      }    }\label{fig2}
\end{figure} 

In the second application, we prove the security of coherent-state QKD. Here, one of the receivers (Alice) is the eavesdropper (renamed to Eve) and her goal is to eavesdrop on the quantum channel connecting the transmitter and the other receiver (Bob). For concreteness, we first consider a phase encoded coherent-state QKD protocol~\cite{Huttner1995}, which uses the encoding scheme $\ket{\psi_{z_0z_1}}$: $\ket{\psi_{00}}=\ket{\sqrt{\mu}}$, $\ket{\psi_{10}}=\ket{-\sqrt{\mu}}$, $\ket{\psi_{01}}=\ket{i \sqrt{\mu}}$, and $\ket{\psi_{11}}=\ket{-i \sqrt{\mu}}$, where $\mu$ is the mean photon number of the coherent state. To maximise the sifting efficiency of the protocol, we use $\{\ket{ \sqrt{\mu}},\ket{-\sqrt{\mu}}\}$ for key generation and $\{\ket{ i \sqrt{\mu}},\ket{-i \sqrt{\mu}}\}$ for testing the security of the channel. Correspondingly on Bob's side, we have that he uses measurement $y=0$ for key recovery and measurement $y=1$ for estimating the channel noise; we write $\eps_0$ and $\eps_1$ to denote the error probabilities observed in the key basis and the test basis, respectively. In this case, the sifting rate of the protocol tends to 1 (in the limit of infinite keys) when the probability of choosing the key basis goes to 1~\cite{Scarani2009}.

\begin{figure}[t!]
\includegraphics[width=8.8cm]{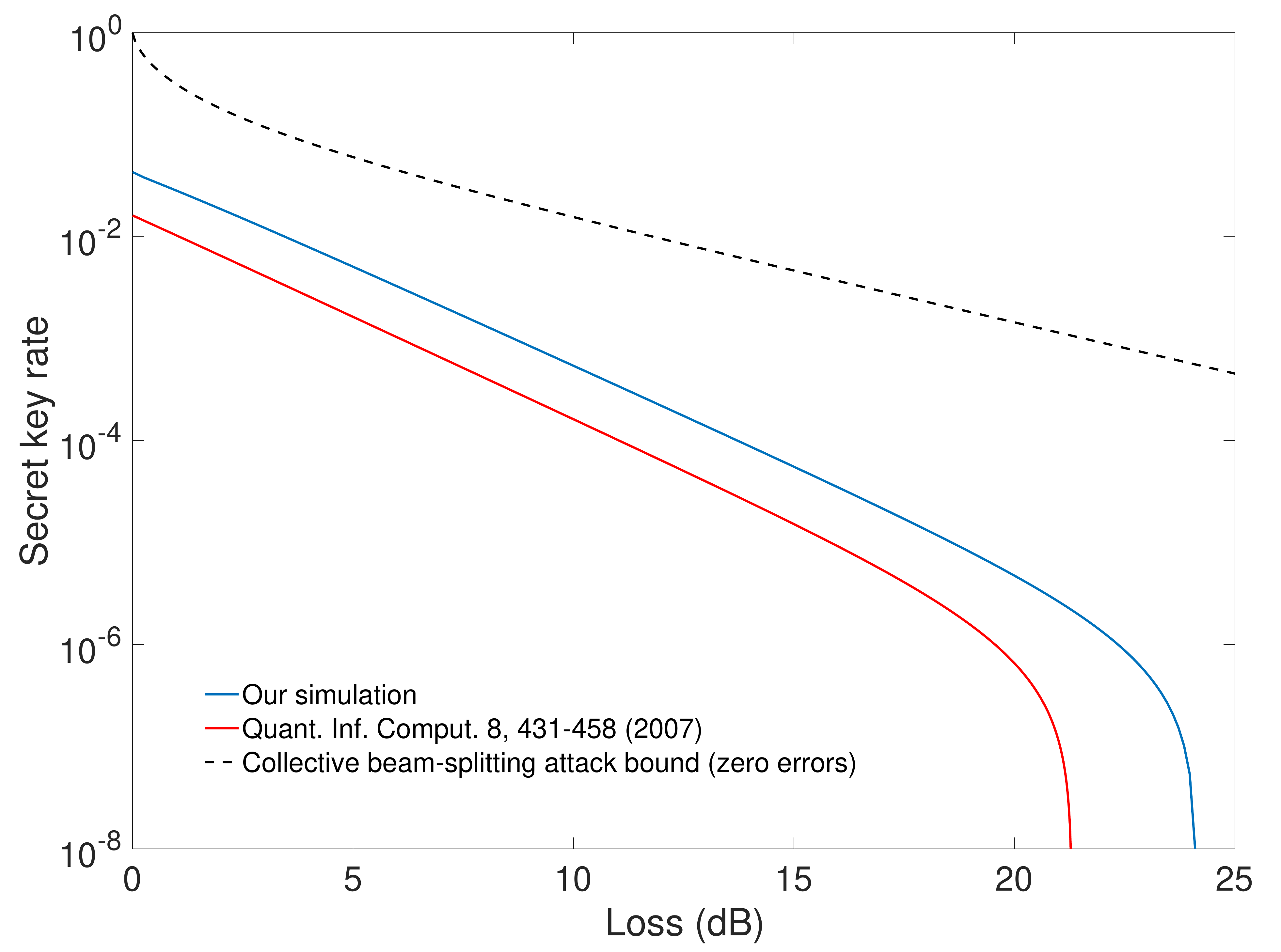}\caption{\footnotesize{\textbf{Phase encoded coherent-state QKD} We compare our secret key rate against the one given in ref.~\cite{Lo2007}. For the key rate simulation, we assume a detector dark count rate of $p_{\rm{dc}}=10^{-7}$ and an intrinsic optical error rate of $2\%$. For a given channel loss $1-\eta$, the probability of detecting a signal is $p_{\rm{det}}=1-(1-p_{\rm{dc}})^2\exp(-2\eta \mu)$ and the error probability is $\eps=(p_{\rm{dc}}+(1-\exp{(-2\mu \eta)})0.02)/p_{\rm{det}}$. Using these, we maximise the expected secret key rate (\ref{eq:keyrate}). More precisely, we perform two optimisations. First, for a given $\mu$ we maximise the phase error rate over $\qset_{2}$ subject to the above constraints. This gives us a lower bound on the achievable secret key rate. Then, we optimise the secret key rate over $\mu$. This gives us an estimate of the optimal secret key rate. Comparing to the secret key rate of ref.~\cite{Lo2007} (red line), we see that our method predicts a higher secret key rate (blue line) for any loss point. For further comparison, we also plotted the collective beam-splitting attack bound\cite{Branciard2007} (the top curve: black dashed line), which serves as an upper-bound on the achievable secret key rate; note that this bound is not tight and it assumes zero errors.    }    }\label{fig3}
\end{figure}

In Appendix~\ref{App:C}, we show that the expected secret key rate (per signal sent) is
\be \label{eq:keyrate}
R_{\rm{key}}^{\infty} \geq  \max \{0, p_{\rm{det}} \left[1-h_2(\eps_0)-h_2(\eps_{\rm{ph}})\right]\},
\ee
where $\eps_{\rm{ph}}$ is the so-called phase error rate of the key basis~\cite{Shor00}, $p_{\rm{det}}$ is the probability of detection, and $h_2(\cdot)$ is the binary entropy function. The quantity of interest here is $\eps_{\rm{ph}}$, which is maximised assuming fixed system parameters (e.g., $\mu$, $\eps_0$ and $\eps_1$). More specifically, we use the second level of the hierarchy $\Sset_{2}$ and maximise $\eps_{\rm{ph}}$ over the set of compatible probabilities in $\qset_{2}$. The outcome of the numerical optimisation is shown in Figure \ref{fig3} along with the simulation parameters. To benchmark our results against the best known security analysis for the protocol, we also plot the security bound of ref \cite{Lo2007} using the same constraints. From the figure, it is evident that our secret key rates are always higher than the ones given by ref.~\cite{Lo2007}. Importantly, this shows that our method significantly improves the security and feasibility of practical QKD, despite making only a few assumptions about the implementation. For completeness, we note that refs.~\cite{Coles2016,Winick2017} have also recently proposed a new security proof technique based on semidefinite programming (but using a completely different approach). In the case of the current protocol, their simulation outcomes are similar to ours, however their method additionally requires that Bob's measurements are fully characterised and an optical squashing model exists for the measurements~\cite{Beaudry2008}.

\begin{figure}[t!]
\includegraphics[width=8.7cm]{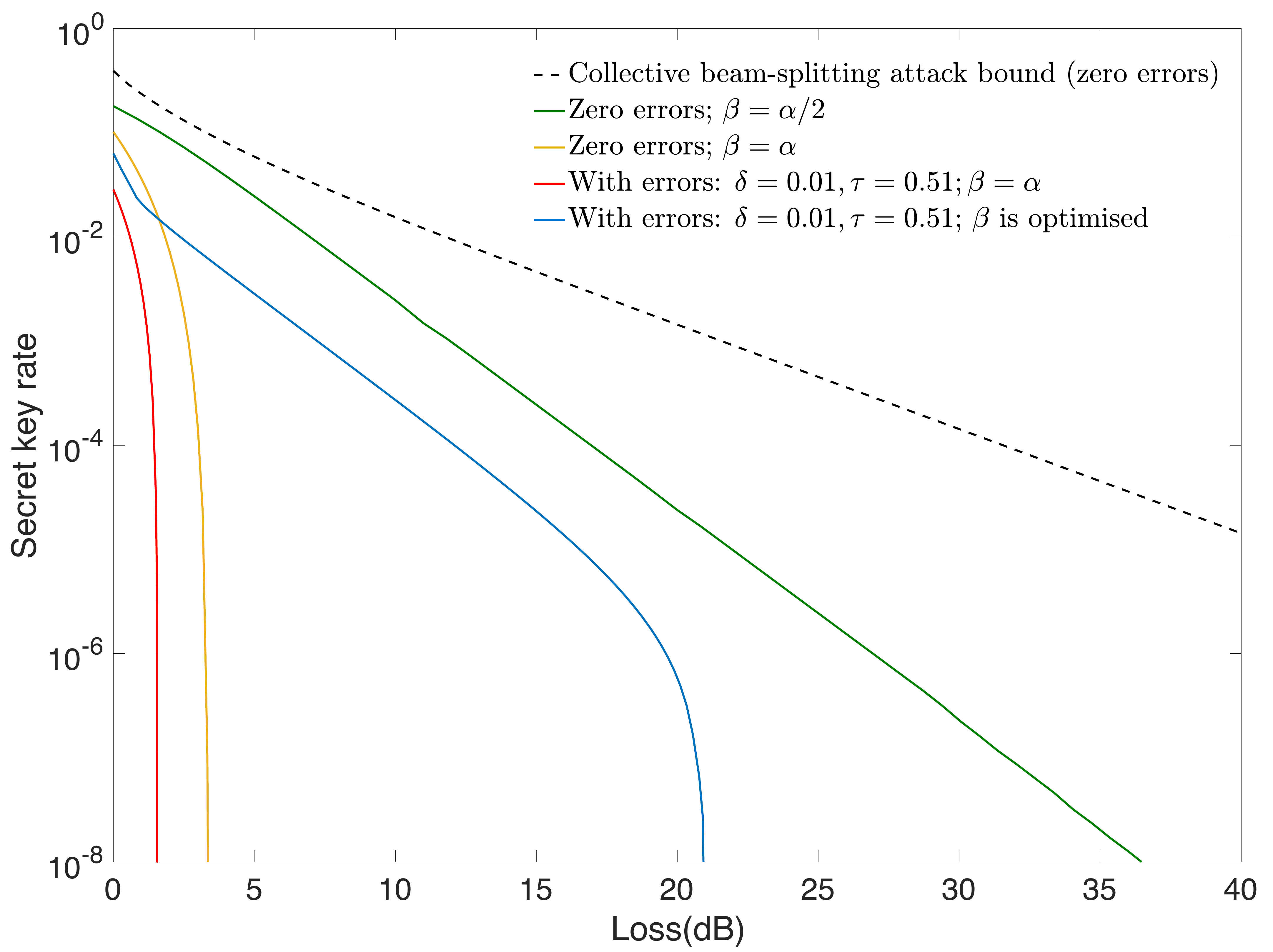}\caption{\footnotesize{\textbf{Time encoded coherent-state QKD} For this simulation, we consider an error model that is based on imperfect intensity modulation and imperfect mixing of coherent states. On the transmitter's side, we assume that the intensity modulator used to perform the on-off keying has finite extinction ratio, i.e., states are prepared as $\ket{\sqrt{1-\delta}\alpha}$ and $\ket{\sqrt{\delta}\alpha}$ instead of $\ket{\alpha}$ and $\ket{0}$. Here, we use $\delta=0.01$. On the receiver side, we assume that the beam-splitter in the measurement scheme of ref.~\cite{Moroder2012} has a ratio of $51/49$ instead of the ideal $50/50$. Using these component models and assuming that each detector has a dark count rate of $p_{\rm{dc}}=10^{-7}$, we run the optimisation as per the previous QKD application and obtain four sets of data points using the original COW QKD encoding and a new encoding strategy where the test sequence is optimised. For each of these, we compute the secret key rates with and without errors.  }}\label{fig4}
\end{figure}

To demonstrate the ability of our method to handle non-standard QKD protocols, we consider the security of a modified coherent-one-way (COW) QKD protocol~\cite{Stucki2005,Korzh2015}, which is based on the transmission of time encoded coherent states $\{\ket{0}\ket{\alpha}, \ket{\alpha}\ket{0}, \ket{\alpha}\ket{\alpha}\}$ with $\alpha=\sqrt{\mu}$. Here, the first two sequences of coherent states carry the secret bit (i.e., `0' $\rightarrow \ket{0}\ket{\alpha}$ and `1' $\rightarrow \ket{\alpha}\ket{0}$) and the last sequence is a test state used to estimate Eve's information about the secret bit. For Bob's measurements, we use the active switching measurement scheme proposed in ref.~\cite{Moroder2012} instead of the original passive switching scheme~\cite{Stucki2005}. In this setup, Bob employs an optical switch to send the incoming states either into the data line or the monitoring line: the former measures the arrival time of the incoming states while the latter measures the coherence (the interference visibility) between two adjacent states. The advantage of this scheme is that it yields higher detection probabilities than the passive scheme. Another major modification is that only the coherence of the test sequence $\ket{\alpha}\ket{\alpha}$ is measured. More specifically, the variant protocol does not measure the coherence between adjacent encodings (e.g., in cases like $\ket{0}\ket{\alpha}; \ket{\alpha}\ket{0}$ or $\ket{\alpha}\ket{\alpha}; \ket{\alpha}\ket{0}$) like in the original protocol. This modification is largely motivated by earlier research which showed that knowing the coherence information between adjacent encodings does not significantly improve the security of the protocol~\cite{Moroder2012}. Importantly, in discarding these events, we have two benefits. Firstly, the security analysis is greatly simplified, i.e., we only need to analyse a single encoding instead of a sequence of encodings, which can be unwieldy. Secondly, this opens up the possibility to explore scenarios whereby the mean photon number of the test sequence $\ket{\alpha}\ket{\alpha}$ is optimised. More concretely, we can now adjust the mean photon number of the test sequence to maximise the secret key rate. In the following, we will use $\ket{\beta}\ket{\beta}$ to represent the optimised test sequence. 

Using the same approach as before (i.e., equation (\ref{eq:keyrate})), we compute the secret key rate of the variant protocol using a realistic error model that assumes an imperfect intensity modulator (on the transmitter side) and an imbalanced beam-splitter on the receiver side; see the description of Figure~\ref{fig4} for more details. We first simulate the expected secret key rates of the protocol using the original COW QKD test sequence $\ket{\beta=\alpha}\ket{\beta=\alpha}$ with errors (red curve) and without errors (yellow curve). Both of these curves show that secret keys can only be distributed in the low loss regime (i.e., less than 4 dB loss; or equivalently 20 km of optical fibre length). Comparing to the collective beam-splitting attack curve~\cite{Branciard2007} (black dashed curve), we observe that the original COW QKD encoding may be sub-optimal. To investigate this possibility, we use the flexibility of our method and further optimise $\ket{\beta}\ket{\beta}$ over a discrete set of ratios $\beta/\alpha$ to search for the best test sequence for a given loss point. We find that the improvement is highly significant. In the case with zero errors, the optimal ratio is $\beta=\alpha/2$ and the tolerable loss is extended to more than 35 dB, which spells a $\geq 30$ dB improvement over the original COW QKD encoding. The secret key rates (green curve) are also significantly higher and are close to the collective beam-splitting attack bound (in the low loss regime). In the case with errors, we also see similar improvements. More concretely, the optimised variant protocol is now able to distribute secret keys up to about 21 dB loss with errors, which translates to a fibre distance of about 110 km. In conclusion, our findings strongly indicate that it is much more secure to vary the mean photon number of the test sequence. 
\newline

\noindent \textbf{Outlook.~}Taken together, our findings thus provide a powerful method to analyse the quantum set of any discretely modulated P\&M quantum network, independently of how the network and decoding measurements are implemented. From the perspective of quantum information theory, the toolbox can help to reveal the fundamental limits of quantum communication and to analyse the performance of any quantum coding scheme. On the application side, the toolbox can be used to analyse the performance of quantum network protocols and the security of quantum cryptography, as evidenced by the three examples given above. Concerning the latter, it would be interesting to investigate how the toolbox could be utilised to solve the other open problems in quantum cryptography. For example, the security of round-robin differential phase-shift QKD~\cite{Sasaki2014} or continuous variable QKD protocols with discrete modulations~\cite{Leverrier2009}.

 \bigskip
\noindent \textbf{Acknowledgements.~}We acknowledge support from the National University of Singapore and the Centre for Quantum Technologies.
\newpage
\begin{widetext}
\appendix

\setcounter{equation}{0}
\def\theequation{A\arabic{equation}}

\section{Convergence of the hierarchy} \label{App:A}

In the main text  we saw how to construct an infinite sequence of decreasing semidefinite programming outer approximations to the quantum set  $\qset$.
In the following, we show that the proposed hierarchy converges to $\qset.$ The proof is a straightforward generalisation of the convergence proof from \cite{Miguel2008}.

For any integer $\ell \ge 1$ let  ${\rm SDP}_{\ell}$ be the semidefinite program of size $n|\Sset_\ell|\times n|\Sset_\ell|$ 
defined by the linear constraints identified in the main text.
Let $\Gmat^{\ell}$ be a feasible solution to ${\rm SDP}_{\ell}$. For the convergence proof, it is instructive to think of  $\Gmat^{\ell}$ as an $n\times n$ block matrix, where each  block has size $|\Sset_\ell| \times |\Sset_\ell|$, i.e.,
\be
\Gmat^{\ell}=\sum_{zz'}G^{\ell, zz'}\otimes \ketbra{e_z}{e_{z'}},
\ee
where the entries of $G^{\ell, zz'}$ are indexed by $\Sset_\ell$. Furthermore, let $\qset_\ell$ be the projection of the set of feasible solutions
to  ${\rm SPD}_{\ell}$ to the relevant coordinates,~i.e.,
$$\qset_\ell=\left\{\left(G^{\ell, zz'}\right)_{z,a,b}: \quad  \gell=\sum_{zz'}G^{\ell, zz'}\otimes \ketbra{e_z}{e_{z'}} \text{ is feasible for } {\rm SDP}_{\ell}\right\}.$$

\medskip
\noindent {\bf Theorem:}
We have that  $\qset=\cap_{\ell=1}^\infty \qset_\ell.$
\begin{proof}
We have already seen  that $\qset\subseteq \cap_{\ell=1}^\infty \qset_\ell.$ We now show the other inclusion.

We start with some comments concerning the proof. First, to ease notation, we assume that measurement  outcomes  uniquely define the measurement they correspond to. We denote by $X(a)$ (resp.~$X(b)$) the measurement corresponding to Alice's outcome $a$  (resp.~Bob's outcome $b$). This convention allows us to  write $p(ab|z)$ as opposed to $p(ab|xy,z)$, which is used in the main text. A quantum realization of $p(ab|z)$ is given by $p(ab|z)=\braz E_aE_b\ketz, $ where $E_a^\dag=E_a$, $E_aE_{a'}=\delta_{a,a'}E_a$  if $X(a)=X(a')$, and $[E_a,E_b]=0$. Second, for any $z,z'\in [n],$ the $|\Sset_\ell| \times |\Sset_\ell|$ matrix $G^{\ell, zz'}$ is indexed by the operators in the set $\Sset_\ell$.  Operators $U , E_a, E_aU,$ and $\mathbb{I}$  will be associated with row or column indices, $u,a, au,$ and $1,$ respectively. Consider a distribution $p(ab|z) \in \cap_{\ell=1}^\infty \qset_\ell$ and let $\gell$ be a level-$\ell$ certificate.

\medskip

\noindent {\bf Step 1:} For any integer $\ell \ge 1$, we have that
\be\label{eq:bounded}
|\gell_{(z,u),(z',v)}|\le  1, \text{ for all  } z\in [n]  \text{ and }  u,v\in\Sset_\ell,
\ee where $G^\ell_{(z,u), (z',v)}$ is the $(u,v)$ entry of $G^{\ell, zz'}$
As  $\gell$ is positive semidefinite, it suffices to show that
\be\label{cdwvcerg}
 0\le \gell_{(z,u), (z,u)}\le 1,  \text{ for all } z\in [n] \text{ and } u\in\Sset_\ell.
 \ee
 To get  \eqref{eq:bounded} from \eqref{cdwvcerg},  consider  the $2\times 2$  principal submatrix of $\gell$ indexed by   $(z,u) $ and $(z',v)$, i.e., $$\begin{pmatrix}\gell_{(z,u), (z,u)}& \gell_{(z,u), (z',v)}\\\gell_{(z',v), (z,u)} & \gell_{(z,v) ,(z,v)} \end{pmatrix}.$$
As this matrix  is PSD, its determinant is nonnegative. Combined with \eqref{cdwvcerg}, this  implies that
$$|\gell_{(z,u), (z',v)}|\le  {\sqrt{\gell_{(z,u), (z,u)}\gell_{(z,v) ,(z,v)}}} \le1.$$
Lastly, we prove \eqref{cdwvcerg}. Trivially, we have that $\mathbb{I}^\da E_a=E_a^\da E_a$, and thus,  $\gell_{(z,1), (z,a)}=\gell_{(z,a), (z,a)}$. As $\gell$ is PSD, the $2\times 2$  principal submatrix of $\gell$ indexed by the words  $(z,1)$ and $  (z,a)$ is also PSD, i.e.,
$$\begin{pmatrix}\gell_{(z,1), (z,1)} & \gell_{(z,1), (z,a)}\\\gell_{(z,a), (z,1)} & \gell_{(z,a), (z,a)} \end{pmatrix} =\begin{pmatrix}1 & \gell_{(z,a), (z,a)} \\\gell_{(z,a), (z,a)}  & \gell_{(z,a), (z,a)}  \end{pmatrix} \succeq0,$$
which in turn implies that $0\le \gell_{(z,a), (z,a)} \le 1$ for all $a$.

\medskip
\noindent {\bf Step 2:} Next we embed all matrices $\gell$  in a single normed space (we need to do this as they have different sizes) where the sequence has  a convergent subsequence. For this, we append zeros to extend each matrix $\gell$ to an infinite matrix $\hat{\gell}$, which is indexed by all words  $|u|,|v|=0,1,2,..$.

Now, by Step 1, all matrices  $\hat{\gell}$ lie  in the unit ball of the Banach space $\ell_\infty$. Furthermore, it is well-known that  $\ell_\infty$ is the dual  space of $\ell_1$ \cite{RS}. Thus, by the Banach-Alaogly theorem \cite{RS}, the sequence $(\hat{\gell})_\ell$ has a convergent subsequence,  with respect to the weak$^*$ topology,~i.e., there exists an infinite matrix ${G}^\infty$ such~that
 \be\label{eq:convergence}
 \hat{{G}}^{\ell_j}\overset{w^*}{\rightarrow}\ginf,  \  \text{ as } j\to \infty.
 \ee
  Equation \eqref{eq:convergence} has two important consequences. First, by the definition of the weak* topology, it also implies point-wise convergence, i.e.,
  \be\label{eq:convergence1}
 \lim_{j\to \infty} \hat{{G}}^{\ell_j}_{(z,u), (z',v)}=\ginf_{(z,u), (z',v)}, \text{ for all } z,z'\in[n] \text{ and words } u,v.
 \ee
Second, the infinite matrix $\ginf$ is a PSD kernel \cite{PSDkernel}.
As $\gell$ is feasible for ${\rm SDP}_\ell$ we  have that 
\be
\gell_{(z,a),(z,b)}=p(ab|z),   \ \gell_{(z,1),(z',1)}=\lambda_{zz'},   \ \gell_{(z,u_1),(z',v_1)}=\gell_{(z,u_2),(z',v_2)}, \text{ when } U_1^\da  V_1=U_2^\da V_2,
 \ee
which combined with  \eqref{eq:convergence1} implies that
 \be\label{csdvdever}
 \ginf_{(z,a),(z,b)}=p(ab|z), \   \ginf_{(z,1),(z',1)}=\lambda_{zz'}, \ \ginf_{(z,u_1),(z',v_1)}=\ginf_{(z,u_2),(z',v_2)}\text{ when } U_1^\da  V_1=U_2^\da V_2.
\ee
Furthermore,  as $\ginf$ is a PSD kernel, there exists a (possibly infinite  dimensional) Hilbert space $(\mathcal{H}, \la\cdot ,\cdot \ra)$ and vectors $\{x_{(z,u)}: z=1,\ldots, n, |u|\in \mathbb{N}\}\subseteq \mathcal{H}$ such that
 \be\label{infinitegram}
 \ginf_{(z,u), (z,v)}=\la x_{(z,u)} ,x_{(z,v)}\ra, \text{ for all } u,v,  \text{ and } z\in [n].
 \ee
 This process is the infinite dimensional analogue of the Cholesky decomposition~\cite{Horn2013}.
 
 \medskip
\noindent  {\bf Step 3:} Using the vectors from \eqref{infinitegram}  we construct a quantum realization for  $p(ab|z) \in \cap_{\ell=1}^\infty \qset_\ell$. Specifically, for $z\in [n]$    define the quantum states  $\ket{\psi_z}=\ket{x_{(z,1)}}$ and the measurement operators
\be
 \quad
 \tE_{a}={\rm proj}({\rm span}(\ket{x_{(z,ua)}}: z\in [n], u\in \mathcal{W}))\quad  \text{ and } \quad \tE_{b}={\rm proj}({\rm span}(\ket{x_{(z,ub)}}: z\in [n], u\in \mathcal{W})).
 \ee
 It remains to show that $ \tE_{a}, \tE_b$ such that $p(a,b|z)=\bra{\psi_z}\tE_a\tE_b\ket{\psi_z}$. From the relation
 \be (E_aU)^\dag E_{a'}V=\delta_{a,a'}U^\dag E_aV,  \text{ when } X(a)=X(a'),
 \ee
 we get that
 \be\label{cever}
 \ginf_{(z,au), (z',a'v)}=\delta_{a,a'}\ginf_{(z,u),(z',av)}, \text{ for all }z,z'\in [n].
 \ee
Using \eqref{infinitegram}, Equation \eqref{cever}   implies  that
 \be\label{ferfgrgr}
 \braket{x_{(z,au)}}{x_{(z',a'v)}}=\delta_{a,a'}\braket{x_{(z,u)}}{x_{(z',av)}},  \text{ when } X(a)=X(a').
 \ee
 In particular, \eqref{ferfgrgr} implies that
 \be
  \braket{x_{(z,au)}}{x_{(z',a'v)}}=0,  \text{ when } X(a)=X(a') \text{ and }  a\ne a',
 \ee
 which implies that
 \be
 \tE_a\tE_{a'}=\delta_{a,a'}\tE_a,   \text{ when } X(a)=X(a').
 \ee
 Furthermore,  we have that
 \be\label{important}
 \tE_a\ket{x_{(z,u)}}=\tE_a\ket{x_{(z,au)}}+\tE_a(\ket{x_{(z,u)}}-\ket{x_{(z,au)}})=\tE_a\ket{x_{(z,au)}}=\ket{x_{(z,au)}},
 \ee
 where for the  last equality we use the definition of $\tilde{E}_a$ and for the second to last  equality  we use that $\tE_a(\ket{x_{(z,u)}}-\ket{x_{(z,au)}})=0$; This follows from the following chain of implications:
 \be
 (E_aU)^\dag V=(E_aU)^\dag E_aV \Longrightarrow \ginf_{(z,au),(z',v)}=\ginf_{(z,au), (z',av)}\Longrightarrow \braket{x_{(z,au)}}{x_{(z,v)}-x_{(z,av)}}=0.
 \ee
 Setting $u=1$,  Equation   \eqref{important} implies  that
 \be\label{eq:final1}
 \tE_a\ket{x_{(z,1)}}=\ket{x_{(z,a)}}, \text{ for all } z=1,\ldots,n.
 \ee
 Likewise,  we get that
 \be\label{eq:final2}
\tE_b\ket{x_{(z,1)}}=\ket{x_{(z,b)}}, \text{ for all } z=1,\ldots,n.
 \ee
 Using induction, it follows from \eqref{eq:final1} and \eqref{eq:final2} that
 \be
 U\ket{x_{(z,1)}}=\ket{x_{(z,u)}},
 \ee
 which in turn implies that
 \be\label{cweverg3}
 \ginf_{(z,u), (z',v)}=\braket{x_{(z,u)}}{x_{(z',v)}}  =\bra{x_{(z,1)}}U^\dag V \ket{x_{(z',1)}}.
 \ee
Lastly, combining \eqref{csdvdever} with \eqref{cweverg3} we get that:
\be
 p(ab|z)   = \ginf_{(z,a),(z,b)}=\bra{\phi_z}\tE_a\tE_b\ket{\phi_z}\quad
   \ee
and furthermore
 \be
 \lambda_{zz'} = \ginf_{(z,1),(z',1)}=\braket{\phi_z}{\phi_{z'}}.
\ee
 The last step of the proof is to show that $[\tE_a,\tE_b]=0$. For this note that
 $(E_aU)^\da E_bV=(E_bU)^\da E_aV$ which implies that
 $$\ginf_{(z,au), (z',bv)}=\ginf_{(z,bu), (z', av)}.$$
 In turn, using \eqref{cweverg3} this
 is equivalent to
{$$\bra{x_{(z,1)}}U^\dag [\tE_a, \tE_b]V\ket{x_{(z',1)}}=0, \text{ for all }z,z'\in [n], U, V,$$}
which implies that $ [\tE_a, \tE_b]=0$.

\end{proof}

\section{Distributed QRAC}\label{App:B}
Here we present the semidefinite program for the distributed QRAC protocol. To start with, let us first present the quantum characterisation problem, which is modelled by a set of projective operators $\{A_x^a\}_{a,x}$ and $\{B_y^b\}_{b,y}$ where $a,b,x,y\in \{0,1\}$ and a set of transformed quantum code states $\{\ket{\phi_{z}}\}_{z}$, where ${z=(z_0,z_1) \in\{0,1\}^2}$. For clarity,  in this section we have changed the notation of the measurement operators from $E^a_x$ to $A^a_x$ to distinguish Alice's measurement operators from Bob's. We assume nothing about these operators and states, except that the Gram matrix of $\{\ket{\phi_{z}}\}_z$ {coincides with the Gram matrix of the vectors} 
$\{\ket{+},\ket{-},\ket{+i},\ket{-i}\}$. More precisely, we require that $\braket{\phi_{z}}{\phi_{z'}}=\lambda_{zz'}$, where
\[
\lambda=
\begin{bmatrix}
    &1&\frac{-i+1}{2}&\frac{-i-1}{2}&0 \\
    & \frac{i+1}{2}&1&0&\frac{i-1}{2} \\
    &\frac{i-1}{2}&0&1&\frac{i+1}{2} \\
    &0&\frac{-i-1}{2}&\frac{-i+1}{2}&1 \\
\end{bmatrix}.
\]
Using this quantum model, {Alice's} and Bob's guessing probabilities are given by
$$
\begin{aligned}
& p(a=z_x)=\frac{1}{8}\sum_{a=z_x}p(a|x,z_0z_1)=\frac{1}{8}\sum_{a=z_x}\bra{\phi_{z_0z_1}} A_{x}^a\ket{\phi_{z_0z_1}}, \\
&p(b=z_y)=\frac{1}{8}\sum_{b=z_y} p(b|y,z_0z_1)=\frac{1}{8}\sum_{b=z_y}\bra{\phi_{z_0z_1}} B_y^b\ket{\phi_{z_0z_1}},\\
\end{aligned}
$$ where the quantum code states and measurements are assumed to be randomly chosen (this can be easily generalised to  an arbitrary distribution). To characterise the quantum behaviour of the protocol, we maximise  {Alice's}  guessing probability given that Bob's guessing probability is set to some fixed value $\tau \in[1/2, 1]$. More specifically, we consider the following {optimisation problem:}
\be \label{S1}
\begin{split}
\mathtt{maximize}:&\;p(a=z_x)\\
\mathtt{subject}~\mathtt{to}:
&\;p(b=z_y)=\tau\\
&\;\braket{\phi_z}{\phi_{z'}}=\lambda_{zz'}, {\forall z,z'}\\
& \sum_a A_x^a=\mathbb{I},~\sum_b B_y^b=\mathbb{I}, \ \forall~x,y\\
& A_x^aA_x^{a'}=\delta_{a,a'}A_x^a,~ B_y^bB_y^{b'}=\delta_{b,b'}B_y^b,\  \forall~a,b,x,y\\
& A_x^a B_y^b= B_y^bA_x^a,\ \forall~a,b,x,y. \\
  \end{split}
\ee
However, as mentioned in the main text, this optimisation problem is computationally intractable. To this end, we consider instead the SDP relaxation  of \eqref{S1} corresponding to the set of operators $\Sset_{1}=\{\mathbb{I}, A_0^0, A_0^1, A_1^0, A_1^1, B_0^0, B_0^1, B_1^0, B_1^1\}$.
Using the label $z=(z_0,z_1)\in \{0,1\}^2$ as a classifier, we can partition any feasible solution $\Gmat$ to the SDP    into $16$  blocks $\{\Gmat^{zz'}\}_{zz'}$,  each having  size $9\times 9$. We index the rows and  columns of $G^{zz'}$ by $0,1,\ldots, 8$. The reader may refer to the following exposition of $\Gmat^{zz'}$ for reference:
\begin{equation*}
\scriptsize{
\arraycolsep=0.4pt\def\arraystretch{2}
\left(\begin{array}{c|cc|cc|cc|cc}
\lambda_{z,z'} &
\langle\phi_{z}| A_0^0|\phi_{z'}\rangle &
\langle\phi_{z}| A_0^1|\phi_{z'}\rangle &
 \langle\phi_{z}| A_1^0|\phi_{z'}\rangle &
 \langle\phi_{z}| A_1^1|\phi_{z'}\rangle
 &  \langle\phi_{z}| B_0^0|\phi_{z'}\rangle
 & \langle\phi_{z}| B_0^1|\phi_{z'}\rangle &
 \langle\phi_{z}| B_1^0|\phi_{z'}\rangle &
 \langle\phi_{z}| B_1^1|\phi_{z'}\rangle \\
\hline

\langle\phi_{z}| A_0^0|\phi_{z'}\rangle& \langle\phi_{z}| A_0^0|\phi_{z'}\rangle & 0 &\langle \phi_{z}|A_0^0A_1^0|\phi_{z'}\rangle & \langle\phi_{z}| A_0^0A_1^1|\phi_{z'}\rangle & \langle \phi_{z}|A_0^0B_0^0|\phi_{z'}\rangle& \langle \phi_{z}|A_0^0B_0^1|\phi_{z'}\rangle&\langle\phi_{z}| A_0^0B_1^0|\phi_{z'}\rangle & \langle\phi_{z}| A_0^0B_1^1|\phi_{z'}\rangle  \\

\langle\phi_{z}| A_0^1|\phi_{z'}\rangle & 0& \langle\phi_{z}| A_0^1|\phi_{z'}\rangle  &\langle\phi_{z}| A_0^1A_1^0|\phi_{z'}\rangle& \langle\phi_{z}| A_0^1A_1^1|\phi_{z'}\rangle& \langle\phi_{z}| A_0^1B_0^0|\phi_{z'}\rangle & \langle \phi_{z}|A_0^1B_0^1|\phi_{z'}\rangle &\langle\phi_{z}| A_0^1B_1^0|\phi_{z'}\rangle & \langle \phi_{z}|A_0^1B_1^1|\phi_{z'}\rangle   \\
  \hline
 \langle\phi_{z}| A_1^0|\phi_{z'}\rangle & \langle\phi_{z}| A_1^0A_0^0|\phi_{z'}\rangle& \langle \phi_{z}|A_1^0A_0^1|\phi_{z'}\rangle  &\langle\phi_{z}| A_1^0|\phi_{z'}\rangle & 0& \langle\phi_{z}| A_1^0B_0^0|\phi_{z'}\rangle & \langle\phi_{z}| A_1^0B_0^1|\phi_{z'}\rangle &\langle \phi_{z}|A_1^0B_1^0|\phi_{z'}\rangle & \langle\phi_{z}| A_1^0B_1^1|\phi_{z'}\rangle   \\

 \langle\phi_{z}| A_1^1|\phi_{z'}\rangle& \langle\phi_{z}| A_1^1A_0^0|\phi_{z'}\rangle& \langle\phi_{z}| A_1^1A_0^1|\phi_{z'}\rangle   &0 & \langle\phi_{z}| A_1^1|\phi_{z'}\rangle & \langle\phi_{z}| A_1^1B_0^0|\phi_{z'}\rangle & \langle\phi_{z}| A_1^1B_0^1|\phi_{z'}\rangle &\langle\phi_{z}| A_1^1B_1^0|\phi_{z'}\rangle & \langle \phi_{z}|A_1^1B_1^1|\phi_{z'}\rangle    \\
\hline

\langle\phi_{z}| B_0^0|\phi_{z'}\rangle& \langle \phi_{z}|A_0^0B_0^0|\phi_{z'}\rangle  & \langle \phi_{z}|A_0^1B_0^0|\phi_{z'}\rangle &\langle \phi_{z}|A_1^0B_0^0|\phi_{z'}\rangle & \langle\phi_{z}| A_1^1B_0^0|\phi_{z'}\rangle & \langle\phi_{z}| B_0^0|\phi_{z'}\rangle  & 0 &\langle \phi_{z}|B_0^0B_1^0|\phi_{z'}\rangle & \langle \phi_{z}|B_0^0B_1^1|\phi_{z'}\rangle    \\

 \langle\phi_{z}| B_0^1|\phi_{z'}\rangle& \langle \phi_{z}|A_0^0B_0^1|\phi_{z'}\rangle  & \langle \phi_{z}|A_0^1B_0^1|\phi_{z'}\rangle &\langle \phi_{z}|A_1^0B_0^1|\phi_{z'}\rangle & \langle \phi_{z}|A_1^1B_0^1|\phi_{z'}\rangle &  0 &\langle \phi_{z}|B_0^1|\phi_{z'}\rangle  &\langle \phi_{z}|B_0^1B_1^0|\phi_{z'}\rangle  & \langle \phi_{z}|B_0^1B_1^1|\phi_{z'}\rangle   \\
\hline

 \langle\phi_{z}| B_1^0|\phi_{z'}\rangle & \langle\phi_{z}| A_0^0B_1^0|\phi_{z'}\rangle  & \langle \phi_{z}|A_0^1B_1^0|\phi_{z'}\rangle  &\langle\phi_{z}| A_1^0B_1^0|\phi_{z'}\rangle  & \langle\phi_{z}|A_1^1B_1^0|\phi_{z'}\rangle  & \langle\phi_{z}| B_1^0B_0^0|\phi_{z'}\rangle & \langle \phi_{z}|B_1^0B_0^1|\phi_{z'}\rangle &\langle \phi_{z}|B_1^0|\phi_{z'}\rangle  & 0    \\

 \langle\phi_{z}| B_1^1|\phi_{z'}\rangle& \langle \phi_{z}|A_0^0B_1^1|\phi_{z'}\rangle  & \langle \phi_{z}|A_0^1B_1^1|\phi_{z'}\rangle& \langle \phi_{z}|A_0^1B_1^1|\phi_{z'}\rangle  & \langle \phi_{z}|A_1^1B_1^1|\phi_{z'}\rangle  & \langle \phi_{z}|B_1^1B_0^0|\phi_{z'}\rangle  & \langle \phi_{z}|B_1^1B_0^1|\phi_{z'}\rangle  &0 & \langle \phi_{z}|B_1^1|\phi_{z'}\rangle    \\
\end{array}\right)
}
\end{equation*}

As explained in the main text, the matrix $\Gmat=\sum_{z,z'}^nG^{zz'}\otimes \ketbra{e_z}{e_{z'}}$, where $ z=(z_0,z_1)\in \{0,1\}^2$, is Hermitian PSD and furthermore, its entries  satisfy certain linear relations corresponding to the  algebraic constraints of the measurement  operators and the Gram matrix of the code states. More specifically, we have that:
\be \label{S2}
\begin{split}
&  \Gmat^{zz'}_{(2i-1, 2i)} =0,
 \;\;i=1,...,4\\
 & \Gmat^{zz'}_{(0,0)}  = \Gmat^{zz'}_{(2i-1, 2i-1)}+\Gmat^{zz'}_{(2i, 2i)}=\lambda_{zz'},\;\;i=1,...,4\\
 & \Gmat^{zz'}_{(i,i)}=\Gmat^{zz'}_{(0,i)},\  i=1, \ldots,8\\
 &\Gmat^{zz'}_{(i,j)}=\Gmat^{zz'}_{(j, i)},\;\;\;\;  i=1,...,4,\; j=5,...,8\\
  & \Gmat^{zz'}_{(i,1)}+\Gmat^{zz'}_{(i, 2)}=\Gmat^{zz'}_{(i,k)}+\Gmat^{zz'}_{(i, k+1)},\;\;i=0,...,8,\;k=3,5,7\\
  \end{split}
\ee
Furthermore,  Alice's  guessing probability is given by:  \be
 p(a=z_x)=
 {1\over 8}\left(G^{00,00}_{(1,1)}+ G^{00,00}_{(3,3)}+G^{01,01}_{(1,1)}+G^{01,01}_{(4,4)}+G^{10,10}_{(2,2)}+G^{10,10}_{(3,3)}+G^{11,11}_{(2,2)}+G^{11,11}_{(4,4)}\right),
 \ee
 and similarly,  Bob's guessing probability is given by:
 \be
 p(b=z_y)={1\over 8}\left( G^{00,00}_{(5,5)}+ G^{00,00}_{(7,7)}+G^{01,01}_{(5,5)}+G^{01,01}_{(8,8)}+G^{10,10}_{(6,6)}+G^{10,10}_{(7,7)}+G^{11,11}_{(6,6)}+G^{11,11}_{(8,8)}\right),
 \ee
 By optimising Alice's  guessing probability $p(a=z_x)$ over all  PSD matrices $ \Gmat=\sum_{z,z'}^nG^{zz'}\otimes  \ketbra{e_z}{e_{z'}}$ satisfying the linear constraints described in \eqref{S2}, and additionally,  satisfying  $p(b=z_y)=\tau$ for  a fixed scalar $ \tau \in [1/2, 1 ]$, we get the plot of $(2p(a=z_x)-1)^2+(2p(b=z_y)-1)^2\leq 1/2$, as illustrated in Figure \ref{fig2}.

\section{Coherent-state QKD}\label{App:C}

Here we analyse the asymptotic security of the phase-encoding coherent-state QKD protocol assuming collective attacks~\cite{Scarani2009}; the extension to coherent attacks  is straightforward using either the post-selection technique~\cite{Christandl2009} or the recently developed entropy accumulation theorem~\cite{Dupuis2016}. The security analysis of the time-encoding QKD protocol is the same. To this end, we first start with a (hypothetical but equivalent) purified state \[\ket{\Phi}=\frac{\ket{+}_A\ket{\phi_0}_{BE}+\ket{-}_A\ket{\phi_1}_{BE}}{\sqrt{2}}, \] which is shared between Alice, Bob and Eve after the transmission. The security of this state, with respect to Alice making the qubit measurement $\{\proj{+},\proj{-}\}$ on her share, is given by the Devetak-Winter's key distillation bound~\cite{Devetak2005} and the entropic uncertainty relation for quantum memories~\cite{Berta2010}:
 \[R_{\rm{key}}^{\infty} \geq p_{\rm{det}}\left[1- H(Y|Y') - H(X|X') \right],\] 
 where $X$ and $Y$ are random variables corresponding to Alice's measurement outcomes obtained from $\{\proj{+},\proj{-}\}$ and $\{\proj{+i},\proj{-i}\}$, respectively, and $X'$ and $Y'$ are random variable associated with Bob's measurement outcomes given $y=0$ and $y=1$, respectively. Note we have also assumed that the measurement operators corresponding to detection loss are the same for both measurement settings, i.e., $B^\emptyset_0=B^\emptyset_1=B^\emptyset$; recall that we are using $\{B_y^b\}_{b,y}$ to denote Bob's measurement operators.. This is to ensure that the probability of detecting a signal is independent of Bob's measurement choice, which is needed to rule out detection side-channel attacks exploiting channel loss~\cite{Lo2007} (for example, see ref.~\cite{Lydersen2010}). The key rate can then be further simplified using the Fano's inequality~\cite{Cover2006}, giving
  \[R_{\rm{key}}^{\infty} \geq p_{\rm{det}}\left[1 - h_2(\eps_{\rm{ph}}) - h_2(\eps_0) \right],\] 
 where \[  \eps_{0}=\frac{ \bra{\Phi}(\proj{+} \otimes B_0^1+\proj{-} \otimes B_0^0)\ket{\Phi}}{ p_{\rm{det} }},\quad \eps_{\rm{ph}}=\frac{ \bra{\Phi}(\proj{+i} \otimes B_1^0+\proj{-i} \otimes B_1^1)\ket{\Phi}}{ p_{\rm{det} }}. \] The problem here is that $p_{\rm{det}}$, $\eps_0$ and $\eps_1$ are experimentally accessible, but the phase error rate $\eps_{\rm{ph}}$ (which is related to Eve's information about $X$) is not. However, the phase error rate cannot be arbitrarily free, for the (hypothetically) prepared states corresponding to Alice measuring $\ket{\Phi}$ in the $\{\proj{+i},\proj{-i}\}$ basis are \emph{close} to what she actually sends, i.e., $\ket{\pm i\alpha}$. Hence, the error rate, $\eps_1$, in the parameter estimation basis should be a good estimate of the phase error rate. In ref.~\cite{Lo2007}, the authors quantified the fidelity between the two preparations using a quantum coin argument and provided an upper bound on $\eps_{\rm{ph}}$ in terms of $p_{\rm{det}}$ and $\eps_1$. However, here we directly maximise $\eps_{\rm{ph}}$ using the second level of the hierarchy under the condition that $p_{\rm{det}}$, $\eps_0$ and $\eps_1$ are fixed to some experimental model (see main text).

\end{widetext}


\begin{thebibliography}{00}


\bibitem{Horodecki2009} Horodecki, R., Horodecki, P., Horodecki, M. \& Horodecki, K. Quantum entanglement. \emph{Rev. Mod. Phys.} \textbf{81}, 865?942 (2009).

\bibitem {Bell1964} Bell, J. S. On the Einstein-Podolsky-Rosen paradox. \emph{Physics}  \textbf{1}, 195--200 (1964).

\bibitem{Wiseman2007}Wiseman, H. M., Jones, S. J.\& Doherty, A. C. Steering, Entanglement, Nonlocality, and the Einstein-Podolsky-Rosen Paradox. \emph{Phys. Rev. Lett.} \textbf{98}, 140402 (2007).

\bibitem{Gisin2007}Gisin, N. \& Thew, R. Quantum communication. \emph{Nat. Photonics} \textbf{1}, 165--171 (2007).

\bibitem{Gisin2002}Gisin, N., Ribordy, G., Tittel, W. \& Zbinden, H. Quantum cryptography. \emph {Rev. Mod. Phys.} \textbf{74}, 145--195 (2002).

\bibitem{Yard2011} Yard, J., Hayden, P. \&  Devetak, I. Quantum broadcast channels. \emph{IEEE Trans. Inf. Theory} \textbf{57}, 7147--7162 (2011).

\bibitem {Hirche2015} Hirche, C. \& Morgan, C. An improved rate region for the classical-quantum broadcast channel. \textit{2015 IEEE International Symposium on Information Theory (ISIT)}, 2782--2786 (2015).

\bibitem{Savov2015} Savov, I. \& Wilde, M. M. Classical codes for quantum broadcast channels. \emph{IEEE Trans. Inf. Theory}. \textbf{61}, 7017--7028 (2015).

\bibitem{Wootters1982}
Wootters, W. K. \& Zurek, W. H. A single quantum cannot be cloned. \emph{Nature} {\textbf{299}}, 802--803 (1982).


\bibitem{Fuchs1996}
Fuchs, C. A. \& Peres, A. Quantum-state disturbance versus information gain: Uncertainty relations for quantum information. \emph{Phys. Rev. A} {\textbf{53}}, 2038 (1996).

\bibitem{Horodecki2005}
Horodecki, M., Horodecki, R, Sen (De), A. \& Sen, U. Common origin of no-cloning and no-deleting principles\;-\;Conservation of information. 	\emph{Found. Phys.} {\textbf{35}}, 2041--2049 (2005).

 \bibitem{Holevo1973} Holevo, A. S. Statistical decision theory for quantum systems \emph{J. Multivariate Anal.} \textbf{3}, 337--394 (1973).
 
 \bibitem{Barnum1996}
Barnum, H., Caves, C. M., Fuchs, C. A., Jozsa, R. \& Schumacher, B. Noncommuting mixed states cannot be broadcast. \emph{Phys. Rev. Lett.} {\textbf{76}}, 2818 (1996)

\bibitem{Barnum2007}
Barnum, H., Barrett, J., Leifer, M. \& Wilce, A. Generalized no-broadcasting theorem. \emph{Phys. Rev. Lett.} {\textbf{99}}, 240501 (2007)

\bibitem {Brunner2014} Brunner, N., Cavalcanti, D., Pironio, S., Scarani, V. \& Wehner, S. Bell nonlocality. \emph{Rev. Mod. Phys.} \textbf{86}, 419--478 (2014).

\bibitem {Tsirelson1987} Tsirel'son, B. S.  Quantum analogues of the Bell inequalities. The case of two spatially separated domains.  \emph{J. Sov. Math.} \textbf {36}, 557--570 (1987).

\bibitem {Landau1988} Landau, L. J. Empirical Two-Point Correlation Functions. \emph{Found. Phys.} \textbf {18}, 449 (1988).

\bibitem {Wehner2006} Wehner, S.  Tsirelson bounds for generalized Clauser-Horne-Shimony-Holt inequalities. \emph{Phys. Rev. A} \textbf {73}, 022110 (2006).

\bibitem{Miguel2007} Navascu\'{e}s, M., Pironio, S. \& Ac\'{i}n, A.  Bounding the set of Quantum Correlations, \emph{Phys. Rev. Lett.} \textbf {98}, 010401 (2007).

\bibitem {Miguel2008} Navascu\'{e}s, M., Pironio, S. \& Ac\'{i}n, A.  A convergent hierarchy of semidefinite programs characterizing the set of quantum correlations, \emph{New J. Phys.} \textbf {10}, 073013 (2008).

\bibitem{SDI2011}   Paw\l{}owski, M., \& Brunner, N. Semi-device-independent security of one-way quantum key distribution, \emph {Physical Review A} \textbf{84}, 010302(R) (2011)

\bibitem{Bowles2014} Bowles, J., Quintino, M. T., \& Brunner, N.  Certifying the Dimension of Classical and Quantum Systems in a Prepare-and-Measure Scenario with Independent Devices. \emph{Phys. Rev. Lett. } {\textbf{112}}, 140407 (2014). 

\bibitem{Lunghi2015} Lunghi, T., Brask, J. B., Lim, C. C. W., Lavigne, Q., Bowles, J., Martin, A., Zbinden, H.,
\& Brunner, N.  Self-Testing Quantum Random Number Generator.  \emph{Phys. Rev. Lett.} {\textbf{ 114}}, 150501 (2015).

\bibitem{Woodhead2015} Woodhead, E. \& Pironio, S. Secrecy in Prepare-and-Measure Clauser-Horne-Shimony-Holt Tests with a Qubit Bound. \emph{Phys. Rev. Lett.}  {\textbf{115}}, 150501 (2015).

\bibitem{Berta2015} Berta, M., Fawzi, O., \& Scholz, V. B. Quantum Bilinear Optimization. \emph{SIAM J. Optim.} {\textbf{26}} , 1529-1564 (2016).

\bibitem{Himbeeck2017} Himbeeck, T. V., Woodhead, E., Cerf, N. J., García-Patrón, R. \& Pironio, S. Semi-device-independent framework based on natural physical assumptions. \emph { Quantum} \textbf{1}, 33 (2017).

\bibitem{Brask2017} Brask, J. B. Martin, A., Esposito, W., Houlmann, R., Bowles, J., Zbinden, H., \&  Brunner, N. Megahertz-Rate Semi-Device-Independent Quantum Random Number Generators Based on Unambiguous State Discrimination. \emph { Phys. Rev. Appl.} {\textbf{7}}, 054018 (2017).

\bibitem {Arrazola2014} Arrazola, J. M.  \& L\"{u}tkenhaus, N. Quantum communication with coherent states and linear optics.  \emph{Phys. Rev. A} \textbf {90}, 042335 (2014).

\bibitem {Wilde2013} Wilde, M. M.  \textit {Quantum Information Theory} (Cambridge Univ. Press, New York, 2013).

\bibitem{Horn2013} Horn, R. A. \& Johnson, C. R. \textit{Matrix Analysis: Characterizations and Properties CH.7} (Cambridge Univ. Press, Cambridge, 2013).

\bibitem{Boyd1996} Vandenberghe, L. \& Boyd, S. Semidefinite programming, \emph{SIAM Rev.} \textbf{38}, 49--95 (1996).

\bibitem{Burgdorf2016} Burgdorf, S., Klep, I. \& Povh, J. \emph{Optimisation of Polynomials in Non-Commutative Variables} (Springer, Switzerland, 2016).

\bibitem {Ambainis1999} Ambainis, A., Nayak, A., Ta-Shma, A. \& Vazirani, U. Dense quantum coding and a lower bound for 1-way quantum automata. \textit {Proceedings of the thirty-first annual ACM symposium on Theory of computing--STOC 99} (1999).

\bibitem {Nayak1999} Nayak, A. Optimal lower bounds for quantum automata and random access codes. \textit{40th Annual Symposium on Foundations of Computer Science (Cat. No.99CB37039)} (1999).

\bibitem{Scarani2009} Scarani, V. et al. The security of practical quantum key distribution. \emph{Rev. Mod. Phys}. \textbf{81}, 1301--1350 (2009).

\bibitem {Lo2014} Lo, H. K., Curty, M. \& Tamaki, K. Secure quantum key distribution. \emph {Nat. Photonics} \textbf {8}, 595--604 (2014).

\bibitem {Wiesner1983} Wiesner, S.  Conjugate coding. \emph{SIGACT News}, \textbf{15},78--88 (1983).

\bibitem {Cerf1998} Cerf, N.,J. Asymmetric quantum cloning in any dimension. \emph {J. Mod. Opt.} \textbf{47}, 187 (2000).

\bibitem{Huttner1995} Huttner, B., Imoto, N., Gisin, N. \& Mor, T. Quantum cryptography with coherent states. \emph{Phys. Rev. A} \textbf{51}, 1863--1869 (1995).

\bibitem {Lo2007} Lo, H. K. \& Preskill, J. Security of quantum key distribution using weak coherent states with nonrandom phases. \emph{Quant. Inf. Comput.} \textbf {8}, 431-458 (2007).

\bibitem {Shor00} Shor, P. W.\& Preskill, J. Simple Proof of Security of the BB84 Quantum Key Distribution Protocol. \emph {Phys. Rev. Lett.} \textbf{85}, 441--444 (2000).

\bibitem{Branciard2007} Branciard, C., Gisin, N.,  L\"{u}tkenhaus, N. \& Scarani, V. Zero-Error Attacks and Detection Statistics in the Coherent One-Way Protocol for Quantum Cryptography. \emph{Quant. Inf. Comput.} \textbf{7}, 639--664 (2007).

\bibitem{Winick2017} Winick, A., L\"{u}tkenhaus, N. \& Coles, P. J., Reliable numerical key rates for quantum key distribution. \emph{Quantum} \textbf{2}, 77 (2018).

\bibitem {Coles2016} Coles, P. J., Metodiev, E. M. \& L\"{u}tkenhaus, N. Approach for unstructured quantum key distribution. \textit{Nat. Commun}. \textbf {7}, 11712 (2016).

\bibitem {Beaudry2008} Beaudry, N. J., Moroder, T.  \& L\"{u}tkenhaus, N. Squashing Models for Optical Measurements in Quantum Communication. \emph{Phys. Rev. Lett.} \textbf {101}, 093601 (2008).

\bibitem{Stucki2005} Stucki, D., Brunner, N., Gisin, N., Scarani, V. \& Zbinden, H. Fast and simple one-way quantum key distribution. \emph{Appl. Phys. Lett}. \textbf{87}, 194108 (2005).

\bibitem{Korzh2015} Korzh, B., Lim, C. C. W., Houlmann, R., Gisin, N., Li, M. J., Nolan, D., Sanguinetti, B., Thew, R.  \& Zbinden, H. Provably secure and practical quantum key distribution over 307 km of optical fibre. \emph{Nat. Photonics} \textbf{9}, 163--168 (2015).

\bibitem{Moroder2012} Moroder, T., Marcos, T., Lim, C. C. W., Thinh, L. P., Zbinden, H.  \&   Gisin, N. Security of Distributed-Phase-Reference Quantum Key Distribution. \emph {Phys. Rev. Lett.} \textbf{109}, 260--501 (2012).

\bibitem{Sasaki2014} Sasaki, T., Yamamoto, Y. \& Koashi, M. Practical quantum key distribution protocol without monitoring signal disturbance. \textit{Nature} \textbf{509}, 475--478 (2014)

\bibitem{Leverrier2009} Leverrier, A. \& Grangier, P. Unconditional Security Proof of Long-Distance Continuous-Variable Quantum Key Distribution with Discrete Modulation.  \emph{Phys. Rev. Lett.} \textbf {102}, 180504 (2009).

\bibitem{RS} Reed, M. \& Simon, B.\emph{ Methods of modern mathematical physics. I. Functional analysis, 2nd Edition} (Academic Press, New York, 1980).

\bibitem{PSDkernel} Bochner S. Hilbert distances and positive definite functions. \emph{Ann. of Math.}
{\bf 42}, 647-656 (1941).

\bibitem {Christandl2009} Christandl, M., K\"{o}nig, R. \& Renner, R.  Postselection Technique for Quantum Channels with Applications to Quantum Cryptography. \emph {Phys. Rev. Lett.} \textbf {102}, 020504 (2009).

\bibitem {Dupuis2016} Dupuis, F., Fawzi, O. \& Renner, R. Entropy accumulation. Available online at https://arxiv.org/abs/1607.01796 (2016).

\bibitem {Devetak2005} Devetak, I. \& Winter, A.   Distillation of secret key and entanglement from quantum states. \emph {Proc. R. Soc. Lond. A} \textbf {461}, 207-235 (2005).

\bibitem{Berta2010} Berta, M., Christandl, M., Colbeck, R., Renes, J. M. \& Renner, R. The uncertainty principle in the presence of quantum memory. \emph {Nat. Phys.} \textbf{6}, 659--662 (2010).

\bibitem{Lydersen2010} Lydersen, L., Wiechers, C., Wittmann, C., Elser, D., Skaar, J. \& Makarov, M. Hacking commercial quantum cryptography systems by tailored bright illumination. \emph {Nat. Photonics} \textbf {4}, 686--689 (2010).

\bibitem{Cover2006}
Cover, T. M. \& Thomas, J. A.  \emph{Elements of Information Theory, 2nd Edition}  (John Wiley \& Sons, New
York, 2006). 

\end{thebibliography}
\end{document}